\documentclass[sigconf]{acmart}
\AtBeginDocument{%
  }

\usepackage{xurl} 
\usepackage{amsmath}
\usepackage{algorithmic}
\usepackage{graphicx}
\usepackage{booktabs}
\usepackage{array}
\usepackage{multirow}
\usepackage{makecell}
\usepackage{textcomp}
\usepackage{xcolor}
\usepackage{pifont}

\usepackage{tcolorbox}

\newtcolorbox{rqbox}{
colback=black!5!white,
colframe=black,
boxrule=0.3pt,
left=2pt,
right=2pt,
top=2pt,
bottom=2pt,
arc=1pt,
before skip=6pt,
after skip=6pt
}

\begin{document}

\title{MCP-SandboxScan: WASM-based Secure Execution and Runtime Analysis for MCP Tools}

\author{Zhuoran Tan}
\affiliation{%
\institution{University of Glasgow}
\country{United Kingdom}
}

\author{Run Hao}
\affiliation{%
\institution{Aarhus University}
\country{Denmark}
}

\author{Jeremy Singer}
\affiliation{%
\institution{University of Glasgow}
\country{United Kingdom}
}

\author{Yutian Tang}
\affiliation{%
\institution{University of Glasgow}
\country{United Kingdom}
}

\author{Christos Anagnostopoulos}
\affiliation{%
\institution{University of Glasgow}
\country{United Kingdom}
}

\begin{abstract}
Tool-augmented Large Language Model (LLM) agents create a new supply-chain surface: Model Context Protocol (MCP) tools are installed like third-party packages, yet their outputs can enter the agent’s reasoning context. This enables confused-deputy risks in which attacker-controlled inputs cause otherwise benign tools to exercise legitimate authority over files, environment variables, or network-facing operations and reflect sensitive or instruction-like content into LLM-visible fields. We present SandScope, an MCP-aware audit framework that combines runtime witness detection with semantic tool profiling. SandScope executes portable tools under WebAssembly System Interface (WASI) or drives unmodified MCP servers over standard input/output (stdio), extracts LLM-visible sinks from tool results and prompt/message fields, and reports auditable source-to-sink witnesses from environment, file, and tool-input sources while separately recording network-intent and egress evidence. Its semantic layer recovers declared capabilities from tools/list metadata and static registrations to characterize attack surface when execution is incomplete. We evaluate SandScope on controlled cross-language subjects, an evasion benchmark, and a 100-repository MCP corpus. SandScope completes shallow dynamic scans for 35 repositories and, through a broader semantic profiling pass, recovers metadata for 1,127 tools across 71 repositories, including 886 tools with security-sensitive declared capabilities. A schema-guided exploration pass over the 35 dynamically scanned repositories re-executes 33 and observes source-to-sink witnesses in 12. These results show that SandScope provides practical, auditable evidence for MCP tool risk through controlled execution, MCP-aware sink extraction, runtime witness reporting, and semantic capability profiling.
\end{abstract}



\maketitle

\section{Introduction}

The MCP \cite{anthropic2024mcp} is an emerging protocol that standardizes how LLM hosts connect to external servers, tools, and data sources. MCP enables LLM hosts to invoke external capabilities beyond the model itself, expanding agent functionality but also introducing new security risks. In deployments, MCP tools inherit broad host privileges such as file, environment, and network access \cite{radosevich2025mcpsafetyauditllms,jing-etal-2025-mcip}. Third-party tool ecosystems therefore create a software supply-chain (SSC) surface: tools are installed, and invoked like packages, but their outputs may be treated as reasoning context by an LLM agent \cite{song2025protocolunveilingattackvectors}.

Existing MCP security tooling largely targets either \emph{static} inspection of tool descriptions, instructions, and source code \cite{sha2025mcpscan}, or \emph{proxy-based} runtime monitoring of MCP traffic for guardrails \cite{beurer-kellner2025introducing-mcp-scan}. While valuable, these approaches do not fully capture behaviors that manifest only \emph{inside} the tool process at runtime. For example, obfuscated filesystem (FS) access or secrets derived from runtime file reads may be invisible to signature- or description-level scans yet surface during execution \cite{8836102}. However, executing third-party tools directly on the host to expose such behaviors is unsafe~\cite{10.1007/978-981-95-4674-9_17}, especially when tools may inherit broad local authority.

The central risk we study is a confused-deputy setting~\cite{Zhao2025ParasitesIT}: attacker-controlled external content or LLM-generated arguments may reach an otherwise benign or trusted-but-risky MCP tool. The tool can then exercise legitimate authority over local files, environment variables, or network-facing operations and return sensitive or instruction-like content through MCP fields visible to the LLM. Thus, although malicious or compromised tools remain a secondary supply-chain concern, SandScope focuses on producing runtime evidence for external-input-to-LLM-visible flows without assuming malicious tool code.

Together, these observations expose a runtime-observability and containment gap: MCP auditors need to observe behaviors inside tool processes, including flows from external inputs to LLM-visible outputs, without granting untrusted tools unconstrained host authority. To address this gap, we propose \textbf{SandScope}, an MCP-aware audit framework for third-party tools. SandScope combines two evidence layers: a runtime witness detector that reports observed source-to-sink exposures in LLM-visible MCP
outputs, and a semantic profiler that recovers declared tool capabilities to characterize attack surface and guide testing when execution is incomplete.
The core contributions are:
\begin{itemize}
    \item We define LLM-visible MCP response fields, including tool-result text, prompt/messages fields, and structured return payloads, as protocol-level security sinks, and formulate external-to-sink exposure as runtime risk for MCP tools.
    \item We build a scenario-driven runtime audit pipeline that executes portable tools under WASI or drives unmodified stdio MCP servers, then links environment, file, and tool-input sources to MCP-visible sinks through auditable black-box witnesses, while separately recording network-intent and egress evidence.
    \item We introduce a complementary semantic profiling layer that recovers declared MCP tool capabilities from protocol-level \texttt{tools/list} metadata and static source registrations, enabling attack-surface characterization when dynamic execution fails and cross-validation of declared egress-related capabilities against runtime network evidence.
    \item We evaluate SandScope on controlled cross-language subjects, an evasion benchmark, and a 100-repository MCP corpus, reporting runtime scan outcomes, transformation limits of lightweight witness detection, real-world execution coverage, latency, semantic capability distribution, and semantic/runtime agreement.
\end{itemize}

To our knowledge, SandScope is the first MCP-specific behavioral audit framework that treats LLM-visible protocol fields as security sinks and reports runtime source-to-sink witnesses across unmodified MCP servers and portable WASI subjects.
The source code has been released at~\footnote{https://anonymous.4open.science/r/MCP-SandboxScan-FFFB} to facilitate reproducibility.

\section{Background \& Threat Model}

Agentic AI systems~\cite{11103638} extend LLMs from passive text generators into interactive decision-making systems. Rather than producing a single response, an agent may iteratively observe context, plan actions, invoke external tools, and incorporate tool results into subsequent reasoning steps. This execution pattern allows agents to access capabilities outside the model itself, such as reading files, querying services, calling APIs, or performing computations.

The MCP provides a standardized interface for connecting such LLM hosts to external tools and data sources. In an MCP-style deployment, the host exposes tools to the agent, the agent selects and invokes tools during its decision loop, and the returned tool outputs may be rendered back into the model context. This architecture improves extensibility, but it also creates a security boundary in which tool executions, external inputs, and LLM-visible outputs interact. We therefore model MCP tool execution explicitly before defining the threat model.

\subsection{MCP Tool Execution Model}

We consider an MCP-style agent that invokes external tools as part of its decision loop. A tool consumes the agent-provided invocation context and a set of ambient inputs such as environment variables, local files, and possibly network-derived content. It then produces structured MCP responses and textual outputs that may be fed back into the agent’s prompt. We model a tool invocation as producing a structured record:
\[
Exec := \langle out, err, code, dur \rangle \leftarrow exec(tool; env, data\_dir)
\]
where $out$ and $err$ are bounded output streams, $code$ is the exit status, and $dur$ is the runtime. Crucially, the agent does not consume $\textsc{Exec}$ in isolation. We treat each run as returning a report
\[
R = \langle Exec, Src, Sink, Flow, Sum \rangle.
\]
$Src$ enumerates external-input candidates, including environment values, mounted file contents, tool-call arguments, and network-intent signals.
$Sink$ enumerates LLM-visible regions such as prompt or messages fields and tool-return slots. $Flow$ links source spans to sink spans through lightweight evidence such as substring or decoded-substring witnesses. $Sum$ summarizes execution, evidence, and attack-surface metadata.
Semantic metadata is handled separately as an attack-surface profile rather than as part of the runtime flow witness.

A key observation is that tool outputs are not merely logs. Many agents render tool results into the next-step prompt, such as “Tool said: ...”. Therefore, a string reflected into an MCP tool result is not only disclosed data; it can become instruction-bearing prompt material that influences later reasoning and tool calls.

\subsection{Threat Model}

\begin{figure*}[!htbp]
    \centering
    \footnotesize
    \captionsetup{justification=centering}
    \includegraphics[scale=0.34]{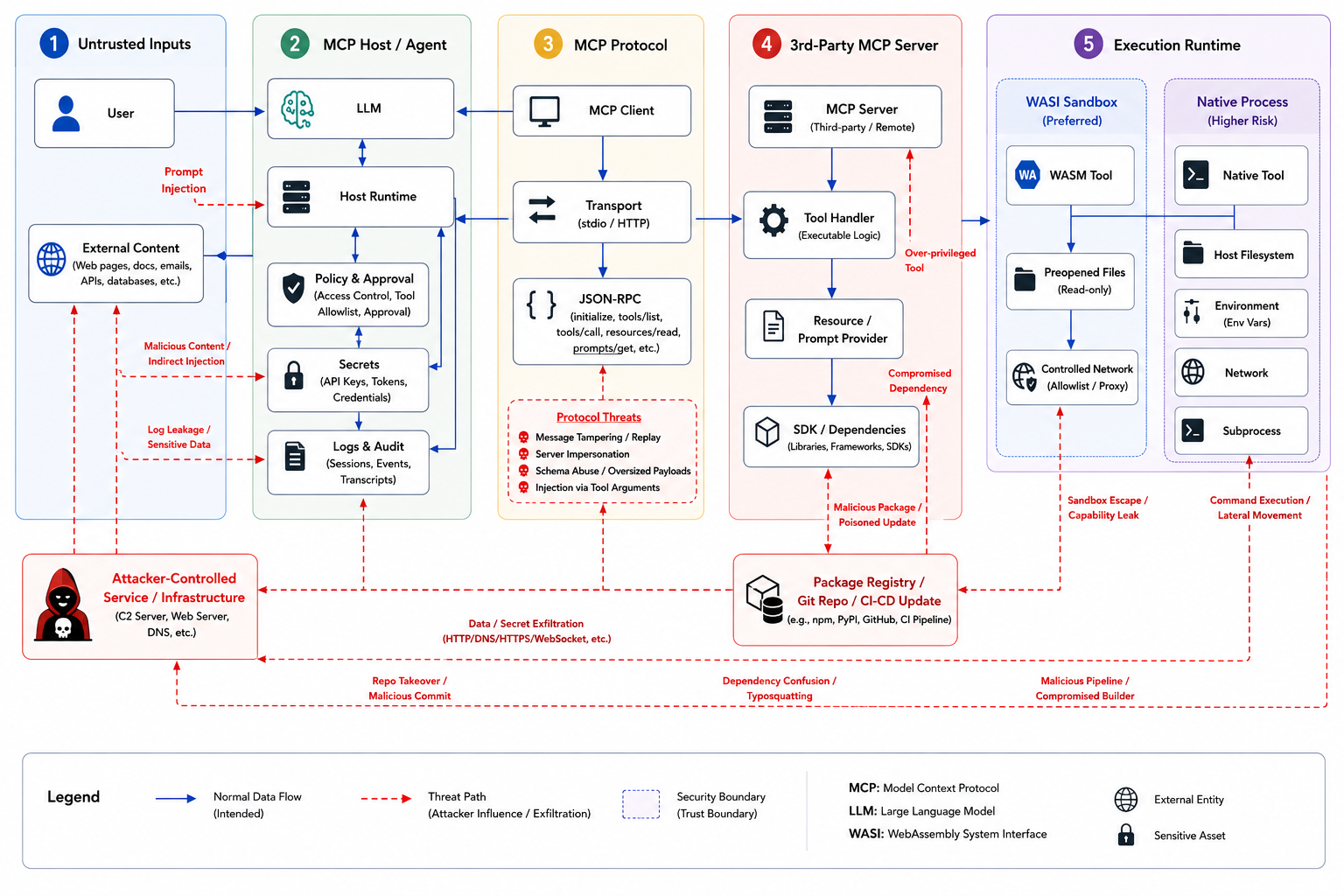}
    \caption{MCP Threat Model - Supply Chain Focus}
    \label{fig:mcp-threat-model}
\end{figure*}

\textbf{Primary adversary}. 
Our primary adversary controls content or arguments that reach an otherwise benign or trusted-but-risky MCP tool. Such inputs include user prompts, LLM-generated tool arguments, retrieved web pages, GitHub issues, README files, API responses, and dependency metadata. The tool itself need not be malicious; the central failure mode is confused-deputy behavior, where attacker-controlled context induces the tool to use its legitimate authority over files, environment variables, or network access and return sensitive data through an LLM-visible MCP response.

\textbf{Secondary adversary}.
Malicious or compromised MCP servers are a secondary supply-chain threat. They motivate contained execution during pre-adoption vetting, but SandScope’s runtime witness detection does not require the tool itself to be malicious.

\textbf{Sources}: We focus on four categories of external inputs that commonly exist in tool execution environments:
\begin{enumerate}
    \item \textbf{Environment sources}: like API keys, user identifiers, system prompts passed via local private environment variables.
    \item \textbf{File sources}: a designated data directory mounted for the tool to read. 
    \item \textbf{Tool-Input sources}: including user-provided, LLM-generated, or schema-generated tool-call arguments.
    \item \textbf{Network-like intents}: strings suggesting outbound requests; in our current prototype this is treated as an observable intent signal rather than full network mediation.
\end{enumerate}
The first three classes can produce byte-level source-to-sink witnesses when their contents appear in MCP-visible sinks, while network-like intents are treated as execution evidence rather than full network-flow mediation.

\textbf{Sinks}: We define sinks as output regions that are likely to be re-inserted into the agent's prompt or otherwise interpreted as instructions. In practice, a sink may be delimited by known markers (e.g., "PROMPT:..."), tool schema conventions, or templated wrapper formats. The central risk we aim to detect is external-to-sink flow: data originating from an external source appears in a sink region.

\textbf{Detection boundary}: SandScope is a positive-evidence detector, not a complete information-flow system~\cite{debenedetti2026defeating}. It reports a witness only when source-derived bytes, or a lightweight normalized/decoded view of them, appear in an extracted LLM-visible sink. Absence of a witness therefore does not prove absence of a flow; Section~\ref{sec:witness-detection} describes the implemented matching strategies and Section~\ref{sec:discussion} discusses limitations.

Figure~\ref{fig:mcp-threat-model} illustrates how these sources and sinks manifest along our execution pipeline, and highlights the corresponding threat points.
We do not claim completeness against sophisticated encoding/obfuscation (e.g., encryption, compression~\cite{7958617}) that breaks string-level matching. We also do not fully mediate arbitrary network behavior in the current prototype; instead, we treat network-like patterns as signals and leave full network syscall instrumentation as future work. 

\subsection{Motivating Example and Scope}

Consider an MCP-enabled coding assistant that installs a third-party GitHub or filesystem server. The server is intended to read repository files, summarize issues, or call external APIs. However, attacker-controlled content in an issue, README, web page, or dependency record may instruct the agent to invoke the tool so that it reads an API token or workspace file and returns it “for debugging.” The tool acts as a confused deputy~\cite{10.1145/3766882.3767177}:
it uses legitimate local or network authority on behalf of untrusted
content, and the result becomes visible to the LLM~\cite{zhang2026mcp}.

SandScope models such third-party content through representative scan scenarios: tool-call arguments, mounted files, environment values, and fixtures that surface explicit fetch or egress intent. The prototype does not crawl the web, fuzz arbitrary arguments, or simulate a full multi-turn agent loop~\cite{NEURIPS2025_5e3661f7}. Separately, semantic profiling recovers declared tool capabilities for attack-surface characterization and prioritization; only runtime witnesses are treated as observed exposure evidence.

\section{SandScope Architecture and Evidence Pipeline}
\label{sec:architecture}

\begin{figure*}[!htbp]
    \centering
    \footnotesize
    \captionsetup{justification=centering}
    \includegraphics[scale=0.355]{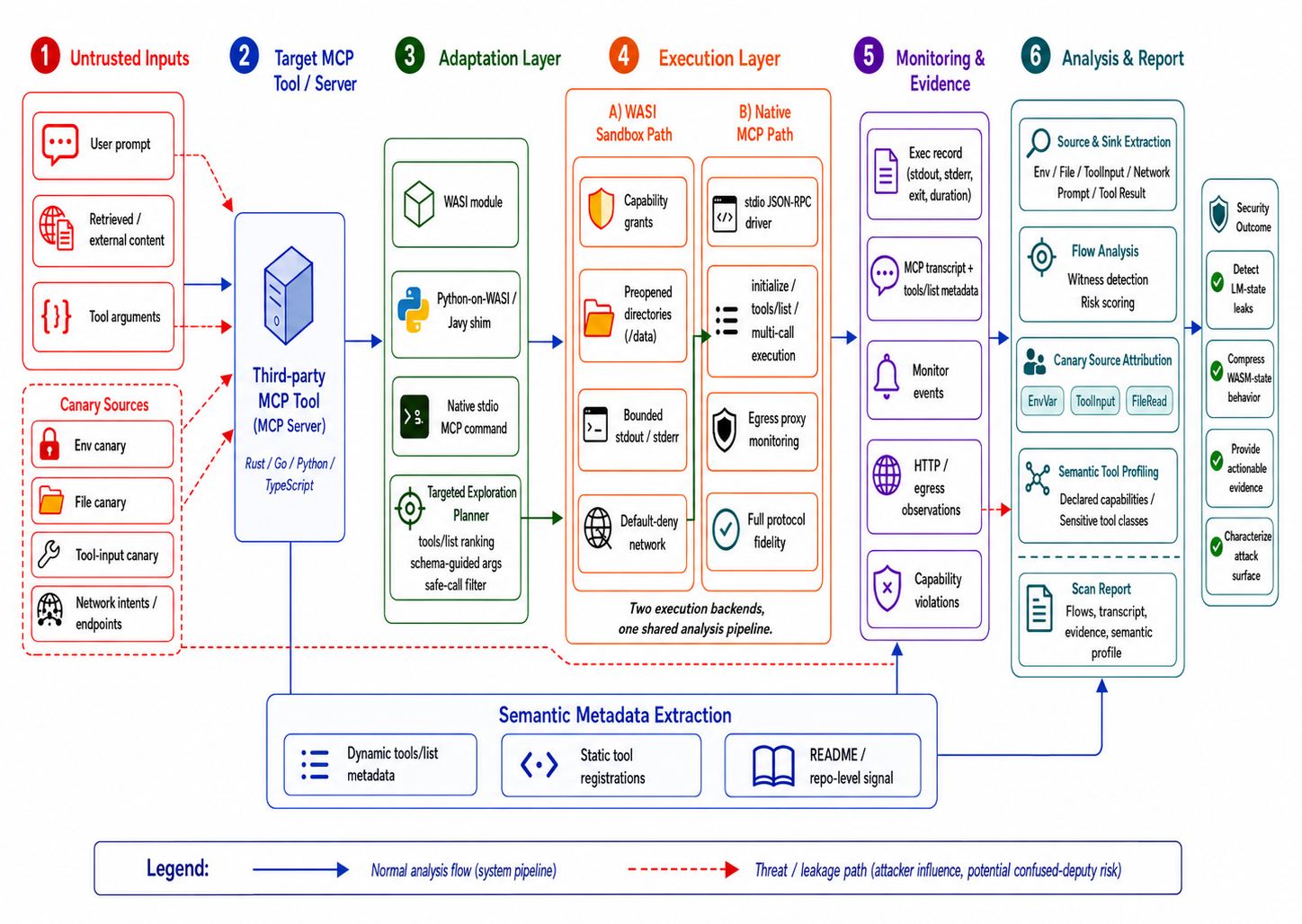}
    \caption{SandScope Framework Overview}
    \caption*{SandScope is a behavioral audit layer for MCP tools}
    \label{fig:system framework}
\end{figure*}

Figure~\ref{fig:system framework} shows SandScope's audit pipeline. SandScope takes an untrusted third-party MCP tool or server, instantiates distinguishable audit sources, adapts the target to either a WASI-backed or native MCP execution path, records boundary-visible runtime evidence, and analyzes MCP-visible outputs for source-to-sink witnesses. In parallel, SandScope extracts semantic metadata from protocol-level \texttt{tools/list} responses and static tool registrations to characterize declared attack surface when execution is incomplete.

SandScope is organized into six stages. First, it seeds environment, file, tool-input, and network-intent canaries that model the sources in our threat model. Second, it identifies the target MCP server or portable tool artifact and selects an adaptation strategy. Third, it executes the target through either a WASI sandbox path or a native standard input/output (stdio) MCP path. Fourth, it records runtime evidence, including execution records, MCP transcripts, monitor events, and egress observations. Fifth, it extracts LLM-visible sinks from prompt-like fields, tool results, and structured return payloads. Finally, it reports source-to-sink witnesses, runtime evidence, semantic capability labels, and scan metadata.

\subsection{Audit Source Setup}
\label{sec:source-instantiation}

SandScope begins by instantiating the external-input sources defined in the threat model. These sources are made distinguishable so that later findings can be attributed to the source category that produced them rather than to a generic output match.

\textbf{Environment sources.}
SandScope seeds selected environment variables with unique canary values. These variables model secrets, tokens, user identifiers, local configuration, and private prompts that may be inherited by an MCP server from the host process.

\textbf{File sources.}
SandScope prepares a designated data directory containing file canaries. In the WASI path, this directory is preopened explicitly. In the native path, it is provided as a controlled workspace fixture. File contents are size-capped before being used as matching sources.

\textbf{Tool-input sources.}
SandScope records user-provided, LLM-generated, or schema-generated tool-call arguments as tool-input sources. During targeted exploration, argument canaries are generated separately from environment and file canaries, allowing SandScope to distinguish direct argument reflection from leakage of ambient authority.

\textbf{Network-intent sources.}
Network-related behavior is handled conservatively. The current prototype does not claim complete network-flow mediation. Instead, it records explicit fetch or egress intents that surface in outputs, MCP responses, runtime evidence, or denied connection events. Thus, environment, file, and tool-input sources can produce byte-level source-to-sink witnesses, whereas network-like behavior is treated as runtime egress evidence unless source-derived bytes are also observed in an LLM-visible sink.

\subsection{Target Discovery and Adaptation Layer}
\label{sec:adaptation-layer}

SandScope supports both portable subjects and unmodified MCP servers. The target discovery stage identifies whether a repository exposes a portable WebAssembly artifact, a tool that can be invoked through a small adapter, or a native MCP server command.

For portable subjects, SandScope uses a WASI-compatible execution path. A tool may enter this path either because it is compiled directly to WebAssembly or because it is invoked through a lightweight adapter that maps the tool's native interface to the expected audit entry point without modifying the tool's core logic. This path is useful for pre-adoption behavioral auditing because the granted capabilities are explicit and reproducible.

For unmodified MCP servers, SandScope uses a native stdio MCP driver. The driver starts the server command, performs protocol initialization, requests \texttt{tools/list}, and invokes one or more \texttt{tools/call} requests while recording the full JSON-RPC transcript. This path preserves protocol fidelity for existing Node.js, Python, Go, Rust, and TypeScript MCP servers.

When \texttt{tools/list} metadata is available, SandScope can run an optional targeted exploration planner. A shallow scan may start a server and recover metadata but still fail to exercise tools that require specific arguments, workspace state, or configuration. The targeted planner therefore ranks candidate tools using names, descriptions, and schemas. It prioritizes non-destructive, high-information tools whose metadata suggests configuration, environment, status, debug, information, listing, reading, workspace, or file-inspection behavior. It filters tools whose names or descriptions suggest destructive or side-effect-heavy operations, such as delete, write, create, update, install, execute, run, or deploy.

For each selected tool, SandScope synthesizes minimal schema-guided arguments. String fields receive a distinguishable tool-input canary; path-like fields receive the path of a mounted file canary; numeric fields receive small constants; booleans receive \texttt{true}; enums use the first declared value; and required object fields are filled recursively. This exploration strategy is intentionally conservative: it improves scan safety and reproducibility but may miss witnesses that require authentication, realistic state, multi-turn interaction, or side-effectful workflows.

\subsection{Controlled Execution Backends}
\label{sec:execution-backends}

SandScope separates MCP-aware analysis from the execution backend used to run the target. It currently supports two execution paths that feed the same monitoring, sink-extraction, and witness-detection pipeline.

\textbf{WASI sandbox path.}
SandScope uses WASI as a containment backend for portable subjects. The WASI backend grants only selected environment variables, preopens a designated data directory, bounds captured standard output and standard error, and denies or mediates network behavior by default. This gives the scanner an explicit capability interface for pre-adoption behavioral auditing.

\textbf{Native stdio MCP path.}
For tools that are not WASI-compatible, SandScope drives unmodified stdio MCP servers and extracts MCP-visible fields from JSON-RPC transcripts. This native path provides protocol fidelity but does not by itself provide the same containment guarantees as WASI. In practice, native scans of untrusted tools should be paired with an OS-level sandbox such as Docker or Bubblewrap~\footnote{https://github.com/containers/bubblewrap}.

SandScope's contribution is therefore not ``WASM instead of OS sandboxing.'' Rather, SandScope provides an MCP-aware behavioral audit pipeline that can use WASI when a portable artifact or adapter is available, and can also drive native MCP servers when preserving the original protocol implementation is necessary.

\subsection{Runtime Monitoring and Evidence Collection}
\label{sec:runtime-monitoring}

During execution, SandScope collects boundary-visible artifacts rather than relying on application-specific instrumentation. These artifacts include bounded standard output and standard error, exit status, elapsed time, MCP transcripts when available, sandbox or monitor events, capability violations, and egress observations.

These artifacts serve two purposes. First, they expose the MCP-visible payloads from which sinks are extracted. Second, they provide execution evidence for actions such as blocked filesystem access, denied connections, or outbound-request attempts. When an egress monitor or OS-level network policy is enabled, SandScope records denied connection attempts as host-and-port observations. Otherwise, network evidence is limited to explicit fetch or egress intents that surface in captured artifacts.

The native MCP driver records protocol-level evidence, including \texttt{initialize}, \texttt{tools/list}, and \texttt{tools/call} messages. A failed individual tool call is recorded in the transcript but does not necessarily fail the repository scan; the driver continues with remaining non-destructive candidates until the call budget or timeout is reached.

\subsection{MCP-visible Sink Extraction}
\label{sec:sink-extraction}

A sink is any textual region extracted from standard output or MCP tool responses that may be incorporated into subsequent prompts by an agent. SandScope focuses on protocol-level and output-level regions that are likely to become LLM-visible. It instantiates sinks in three forms.

\textbf{Marker-based prompt lines.}
If a trimmed output line contains a prompt marker such as \texttt{PROMPT:}, the line is recorded as a sink. This captures tools that explicitly print prompt material for debugging, logging, or downstream handling.

\textbf{JSON prompt and messages fields.}
If an output line or MCP response parses as JSON, the extractor looks for common prompt encodings such as a string-valued \texttt{prompt} field or an array-valued \texttt{messages} field. Message content is concatenated and recorded as a sink when non-empty.

\textbf{Structured tool-return leaves.}
For structured MCP tool results and JSON outputs that do not contain prompt or messages fields, the extractor collects string-valued leaf nodes and records them as tool-return sinks, retaining their JSONPath for reporting. Simple filters exclude metadata-like paths such as status or identifier fields and very short values.

Outputs that do not match any extraction rule are ignored. As a result, SandScope focuses on sinks explicitly surfaced in captured outputs or MCP transcripts. This is sufficient for the behaviors studied in our evaluation, but it does not cover sinks that never appear in observed artifacts.

\subsection{Source-to-Sink Witness Detection}
\label{sec:witness-detection}

For each source string, SandScope generates candidate snippets consisting of the full source string when feasible, fixed-length prefix windows, fixed-length suffix windows, and an optional mid-window for long sources. Suffix windows are included because sensitive values such as API keys, session tokens, and credentials are often partially exposed through logs, debugging output, error messages, or diagnostic channels. Prefix, suffix, and mid-window lengths are heuristic sampling choices that trade recall against matching cost and false positives from short, low-entropy fragments.

For each sink, SandScope matches against the original text, a separator-normalized form, and decoded forms for candidate base64, hex, and ROT13 tokens~\cite{zhao2026proactivedefensellmjailbreak}. This captures common low-effort leaks where source values reach LLM-visible fields after lightweight encoding or chunking.

Each witness records the matched source and sink, snippet, confidence label, and matching strategy, enabling auditability and attribution to exact, normalized, decoded, or source-normalized detector components.

SandScope is intentionally a positive-evidence detector rather than a complete taint-tracking engine. A reported witness means that SandScope observed recoverable source-derived bytes in an LLM-visible sink. The absence of a witness does not prove the absence of a flow. SandScope does not infer arbitrary transformations, encrypted payloads, compressed payloads, hashes, semantic rewrites, summarized content, or flows where no recoverable source-derived bytes are exposed.

\subsection{Semantic Metadata Extraction and Capability Profiling}
\label{sec:semantic-profiling}

Semantic profiling complements runtime witness detection by recovering declared MCP tool capabilities without executing tool calls or assigning vulnerability labels. It provides attack-surface metadata for characterizing repositories when dynamic execution is incomplete and for prioritizing future testing.

SandScope extracts semantic metadata from three sources. First, when the runtime path successfully starts an MCP server, SandScope records the protocol-level \texttt{tools/list} response, including tool names, descriptions, and schemas. Second, when dynamic startup fails, SandScope falls back to static source-level MCP tool registrations, such as tool registration calls, FastMCP decorators, \texttt{server.tool} definitions, or equivalent framework-specific declarations. Third, when no tool registration can be recovered, SandScope records weak repository-level context from README files and repository metadata. These repository-level signals are used only as context and are excluded from tool-level counts.

Each recovered tool is assigned zero or more capability labels. The current taxonomy includes network access, browser automation, cloud or SaaS operation, filesystem access, code repository operation, database access, credential or secret handling, shell or process execution, and unknown or benign utility. Labels are multi-label. A tool is counted as exposing a security-sensitive declared capability if at least one label belongs to a class that can read, write, transmit, or act on external authority.

For repositories that complete dynamic scanning, SandScope cross-validates declared egress risk against runtime egress evidence. Declared egress risk is recorded when at least one recovered tool has a capability that may communicate with external services. Observed egress evidence is recorded when the runtime scan surfaces an explicit network or fetch intent, an egress observation, or an MCP-visible outbound-request attempt. This comparison is performed only over dynamically scanned repositories and is used to measure agreement between declared attack surface and behavior exercised by the scan scenario, not to establish vulnerability ground truth.

\subsection{Scan Report}
\label{sec:scan-report}

SandScope's final report separates three evidence types. Source-to-sink witnesses are runtime findings showing that seeded environment, file, or tool-input bytes reached an LLM-visible sink. Egress observations are runtime evidence of network-like behavior, denied connections, or explicit outbound-request intent. Semantic capability labels are declared attack-surface metadata recovered from protocol-level or static sources.

The report includes the execution outcome, bounded output summaries, MCP transcript references, extracted sinks, witness records, egress observations, semantic capability labels, and scan metadata such as timeout or build failures. This separation is central to SandScope's interpretation: a source-to-sink witness is positive runtime evidence for the exercised scenario, an egress observation is behavioral evidence, and a semantic label is a declared capability rather than a vulnerability claim.

\section{Evaluation}

We evaluate SandScope along four questions:
\begin{itemize}
    \item \textbf{RQ1:} Can SandScope produce auditable runtime witnesses when authority-bearing sources are reflected into LLM-visible MCP sinks under controlled MCP executions? 
    \item \textbf{RQ2:} How does the lightweight witness detector behave under low-effort transformations and known evasion patterns? 
    \item \textbf{RQ3:} What fraction of real-world MCP repositories can be dynamically executed through a shallow corpus scan, and what are the dominant failure modes? 
    \item \textbf{RQ4:} Does targeted schema-guided exploration uncover source-to-sink witnesses in real MCP servers that already complete a shallow protocol session? 
    \item \textbf{RQ5:} How much corpus visibility does semantic profiling provide beyond dynamic execution, and how do declared egress capabilities align with runtime egress evidence?
\end{itemize}

We evaluate SandScope in three settings. First, a controlled cross-language benchmark includes 30 Go, Python, Rust, and TypeScript subjects spanning benign tools, environment leaks, file-exfiltration fixtures, upstream-content echoes, C2-beacon fixtures, and MCP protocol variants; it measures source-to-sink witnesses, false positives and negatives, and execution-path support. Second, a shallow scan of 100 real-world MCP repositories measures dynamic coverage, failure modes, latency, semantic extraction coverage, declared capabilities, and semantic/runtime egress agreement. Third, a targeted exploration pass over shallow-scan successes measures whether schema-guided execution can surface source-to-sink witnesses in real MCP servers.

\subsection{Experimental Setup}

\textbf{Controlled subjects.} The controlled benchmark contains 30 subjects across four implementation ecosystems. Some subjects are direct WASM artifacts, some use a WASM shim, and others are exercised through the native stdio MCP path. Each subject is labeled with an expected outcome: no finding, runtime flow, network-intent signal, or flow plus intent.

\textbf{Real-world corpus.} The real-world corpus contains 100 popular MCP-related repositories. SandScope resolves 91 repositories for dynamic scanning and attempts startup, protocol initialization, \texttt{tools/list} discovery, and tool-call execution. Repositories that fail dynamic execution remain eligible for semantic profiling through static source-level registrations and repository metadata.

\textbf{Targeted exploration subset}. Source-to-sink witnesses require a working MCP protocol session, so the targeted pass is run over the repositories that complete the shallow scan. This pass does not attempt to improve build or startup coverage. Instead, it increases execution depth by ranking tools from \texttt{tools/list}, generating schema-guided non-destructive calls, seeding separate environment, file, and tool-input canaries, and scanning text and structured MCP result leaves for canary witnesses.

\textbf{Execution paths.} The controlled benchmark exercises direct WASM, WASM-shim, and native stdio MCP paths. The real-world corpus primarily uses the native MCP path for unmodified servers, while WASI is used when a portable artifact or shim is available.

\textbf{Sources and sinks.} The evaluation uses the runtime source instantiation and sink extraction rules from Sections~\ref{sec:runtime observation} and~\ref{sec:sink extract}. A flow is emitted only when SandScope observes a recoverable source-derived snippet in an extracted MCP-visible sink.

\textbf{Metrics.} For the controlled benchmark, we report TP, FP, FN, TN, precision, recall, and F1 against expected findings. For the evasion benchmark, we compare raw substring matching with the enhanced detector. For the shallow real-world corpus scan, we report dynamic scan coverage, build/startup/call-timeout failures, ecosystem coverage, WASM compatibility classes, and latency. For the targeted exploration pass, we report input repositories, successful re-executions, flow-observed repositories, total source-to-sink witnesses, and failures. For semantic profiling, we report repositories with tool-level metadata, extracted tools, described tools, tools with at least one security-sensitive declared capability, multi-label capability distribution, and semantic/runtime egress agreement.

\subsection{Controlled Runtime Results}

\begin{table*}[!htbp]
\centering
\small
\setlength{\tabcolsep}{5pt}
\caption{Controlled runtime results by ecosystem. Execution paths are reported as Direct-Wasm / Wasm-Shim / Native-only}
\label{tab:runtime_overview}
\begin{tabular}{lcccccl}
\toprule
\textbf{Ecosystem} &
\makecell{\textbf{No.}\ \textbf{Subjects}} &
\makecell{\textbf{MCP} /\ \textbf{Non-MCP}} &
\makecell{\textbf{Execution}\ \textbf{Paths}} &
\makecell{\textbf{Outcome}\ \textbf{TP/FN/FP/TN}} &
\makecell{\textbf{Detected}\ \textbf{Findings}} &
\makecell{\textbf{Missed}\ \textbf{Cases}} \\
\midrule
Go         & 8 & 4 / 4 & 4 / 0 / 4 & 4 / 1 / 0 / 3 & Flow, Intent & c2-beacon \\
Python     & 8 & 5 / 3 & 0 / 3 / 5 & 4 / 0 / 0 / 4 & Flow, Intent & -- \\
Rust       & 6 & 2 / 4 & 4 / 0 / 2 & 3 / 1 / 0 / 2 & Flow, Intent & c2-beacon \\
TypeScript & 8 & 4 / 4 & 4 / 0 / 4 & 5 / 0 / 0 / 3 & Flow, Intent & -- \\
\midrule
\textbf{Total} & \textbf{30} & \textbf{15 / 15} & \textbf{12 / 3 / 15} &
\textbf{16 / 2 / 0 / 12} & \textbf{Flow, Intent} & \textbf{2 cases} \\
\midrule
\multicolumn{7}{l}{
\textbf{Overall:} Precision = 1.000, Recall = 0.889, F1 = 0.941.
} \\
\bottomrule
\end{tabular}
\end{table*}

Table~\ref{tab:runtime_overview} summarizes the controlled runtime evaluation. The true positives correspond to cases where environment values, file contents, or intent-containing strings are reflected into stdout-derived or MCP-derived sinks. The true negatives correspond to benign echo or upstream-count subjects where no authority-containing source reaches an LLM-visible sink.

The two false negatives occur in non-protocol C2-beacon subjects where the expected finding is an intent signal, but the emitted output does not produce a source-to-sink witness under the current extraction rules. These cases illustrate SandScope's boundary between intent observation and flow evidence: it reports observed source-to-sink flows and explicitly surfaced intents, but does not infer hidden network behavior without protocol evidence, output-surfaced intent markers, or network mediation. The same analysis pipeline is exercised across direct Wasm, Wasm-shim, and native-only paths, separating containment and protocol-fidelity choices from MCP-specific sink extraction and witness detection.

\begin{rqbox}
\noindent\textbf{Answer to RQ1.}
SandScope produces auditable runtime witnesses under controlled MCP executions. Across 30 cross-language subjects, it reports 16 true positives and 12 true negatives with no false positives, achieving precision of 1.000, recall of 0.889, and F1 of 0.941.
\end{rqbox}

\subsection{Evasion Benchmark}

\begin{table}[!htbp]
\centering
\small
\caption{Detection results under different transforms.}
\begin{tabular}{llll}
\hline
\textbf{Transform} &
\makecell{\textbf{Raw-only}\\\textbf{detected}} &
\makecell{\textbf{Enhanced}\\\textbf{detected}} &
\makecell{\textbf{Enhanced}\\\textbf{strategy}} \\
\hline
plain         & yes & yes & exact-substring \\
prefix+suffix & yes & yes & exact-substring \\
suffix-only   & yes & yes & exact-substring \\
rot13         & no  & yes & rot13-decoded \\
hex           & no  & yes & hex-decoded \\
base64        & no  & yes & base64-decoded \\
chunked       & no  & yes & separator-normalized \\
\hline
\end{tabular}
\label{tab:detection-results}
\end{table}

Table~\ref{tab:detection-results} compares raw substring matching with the enhanced witness detector. Raw matching covers only verbatim and simple prefix/suffix exposures, whereas the enhanced detector adds normalized and decoded sink views for low-effort transformations such as ROT13, hex, base64, and separator chunking.

This result is not a general evasion defense. The enhanced detector covers only recoverable transformations included in the benchmark; encrypted, compressed, hashed, semantically rewritten, summarized, or custom-transformed leaks remain outside SandScope's guarantee. A match therefore provides positive runtime evidence, while a non-match does not prove absence of flow.

\begin{rqbox}
\noindent\textbf{Answer to RQ2.}
SandScope's enhanced witness detector captures several low-effort transformations that raw substring matching misses, including ROT13, hex, base64, and separator-chunked disclosures. However, it remains a positive-evidence detector and does not claim completeness against encryption, compression, hashing, or semantic rewriting.
\end{rqbox}

\subsection{Real-World MCP Servers Scanning}
We construct a 100-repository MCP corpus from star-ranked GitHub search results using MCP-specific name, topic, and language queries. After deduplication, we filter out SDKs, documentation, registries, curated lists, and client libraries, and retain the top 100 repositories as a popularity-weighted snapshot of visible MCP server implementations.
This analysis has three goals: measuring which repositories can be dynamically exercised, quantifying how much semantic metadata can still be recovered when dynamic execution fails, and comparing declared egress-related capabilities against runtime egress evidence. Together, these results characterize both the practical applicability boundary of the current prototype and the additional visibility provided by semantic profiling.

\subsubsection{Dynamic execution coverage}

\begin{table}[!htbp]
\centering
\small
\caption{End-to-end coverage and failure breakdown on real-world MCP repositories}
\label{tab:realworld-coverage}
\begin{tabular}{@{}lrrrrrr@{}}
\toprule
Subset & Resolved & Scanned & Success & Build & Startup & \makecell{Call/\\Timeout} \\
\midrule
All & 91 & 35 & 38.5\% & 19 & 31 & 6 \\
T1 & 42 & 28 & 66.7\% & 4 & 10 & 0 \\
T2 & 49 & 7 & 14.3\% & 15 & 21 & 6 \\
\bottomrule
\end{tabular}
\end{table}

Table~\ref{tab:realworld-coverage} reports dynamic scan coverage and failure modes. Coverage is substantially higher for dedicated MCP servers than for broader MCP-related repositories, because Tier-1 servers usually expose clearer entry points, whereas Tier-2 repositories often contain examples, libraries, partial integrations, or MCP-adjacent code that is harder to start automatically.

Most incomplete scans are due to operational reproducibility rather than the witness detector itself. Build failures, startup or protocol-initialization failures, and call-timeout failures reflect packaging conventions, runtime dependencies, server entry-point discovery, credential requirements, and protocol behavior.

\begin{table}[!htbp]
\small
\centering
\caption{Dynamic coverage by implementation ecosystem.}
\label{tab:realworld-ecosystem}
\begin{tabular}{lrrrr}
\toprule
Ecosystem & Resolved & Scanned & Coverage & Failures \\
\midrule
npm & 54 & 25 & 46.3\% & 29 \\
Python & 24 & 8 & 33.3\% & 16 \\
Go & 11 & 2 & 18.2\% & 9 \\
Rust & 2 & 0 & 0.0\% & 2 \\
\bottomrule
\end{tabular}
\end{table}

Table~\ref{tab:realworld-ecosystem} shows that coverage also varies across ecosystems. npm-based servers account for the largest share of resolved repositories and successful scans, while Python and Go show lower coverage due to dependency, startup, and packaging variation. Rust is underrepresented in this corpus, so its zero successful scans should be read as a sample limitation rather than a general claim about Rust MCP servers.

The WASM-compatibility breakdown is an applicability observation rather than a measure of WASI effectiveness. Many wasm-hard repositories are still scanned through the native MCP path, while wasm-ready repositories can still fail because of build scripts, startup behavior, protocol initialization, or tool-call requirements. This supports SandScope's design choice: WASI is useful when a portable artifact or adapter is available, but broad MCP coverage also requires native protocol driving and, for strong containment, OS-level sandboxing.

Among the 35 successful shallow scans, SandScope has a median end-to-end latency of 3.2s and p95 of 24.2s; the long tail is dominated by build and startup variability rather than witness detection.

\begin{rqbox}
\noindent\textbf{Answer to RQ3.}
SandScope dynamically executes a meaningful but incomplete portion of real-world MCP repositories. It resolves 91 repositories and completes shallow scans for 35, with failures mainly caused by build, startup, protocol-initialization, and timeout issues rather than the witness detector itself.
\end{rqbox}

\subsubsection{Targeted source-to-sink exploration}

The shallow corpus scan primarily measures build/startup coverage, protocol observability, and semantic visibility; it is not an exhaustive search for source-to-sink witnesses. Many real MCP tools require specific arguments, credentials, workspace state, or configuration before authority-bearing data is read and reflected. We therefore rerun targeted exploration on repositories that complete the shallow scan.

Table~\ref{tab:targeted-exploration} summarizes the targeted pass. The additional schema-guided calls surface witnesses in a non-trivial subset of executable repositories, but these witnesses should not be interpreted as verified vulnerabilities. They are positive runtime evidence that seeded environment, file, or tool-input bytes reached an LLM-visible MCP sink in the exercised scenario.

\begin{table}[!htbp]
  \centering
  \small
  \caption{Targeted source-to-sink exploration over repositories that completed the shallow scan. }
  \label{tab:targeted-exploration}
  \begin{tabular}{>{\raggedright\arraybackslash}p{1.3cm}rrrrrr}
    \toprule
    Subset & Input & Scanned & Success & \shortstack{Flow\\repos} & Rate & Flows \\
    \midrule
    All shallow-success & 35 & 33 & 94.3\%  & 12 & 36.4\% & 60 \\
    Tier-1 servers      & 28 & 26 & 92.9\%  & 9  & 34.6\% & 55 \\
    Tier-2 related      & 7  & 7  & 100.0\% & 3  & 42.9\% & 5  \\
    \bottomrule
  \end{tabular}
\end{table}

The targeted pass changes the interpretation of the real-world study. The shallow pass quantifies applicability and reproducibility limits, while the targeted pass asks whether deeper non-destructive calls can expose runtime evidence once a working MCP session exists. Remaining no-flow runs mean only that no witness was observed under this exploration budget.

\begin{rqbox}
\noindent\textbf{Answer to RQ4.}
Targeted schema-guided exploration uncovers source-to-sink witnesses in real MCP servers that already complete shallow scanning. Among 35 shallow-success repositories, SandScope successfully re-executes 33 and observes 60 witnesses across 12 repositories, showing that deeper non-destructive tool calls expose additional runtime evidence.
\end{rqbox}

\subsubsection{Semantic profiling coverage}

\begin{table}[!htbp]
\centering
\small
\caption{Semantic extraction coverage}
\label{tab:semantic-full-coverage}
\begin{tabular}{lrrrr}
\toprule
Source & Repos & Tools & Described & Sensitive \\
\midrule
Dynamic \texttt{tools/list} & 35 & 539 & 539 & 418 \\
Static source registration & 36 & 588 & 377 & 468 \\
Repo-level signal only & 27 & 0 & 0 & 0 \\
No semantic signal & 2 & 0 & 0 & 0 \\
\midrule
Total tool metadata & 71 & 1127 & 916 & 886 (78.6\%) \\
\bottomrule
\end{tabular}
\end{table}

We additionally run a semantic-only pass over the full corpus. Unlike runtime witness detection, this pass does not execute tool calls and is used for attack-surface characterization rather than vulnerability detection. It reuses protocol-level \texttt{tools/list} metadata when available, falls back to static source-level MCP tool registrations when dynamic execution fails, and records README or repository-level signals only as weak context.

Table~\ref{tab:semantic-full-coverage} shows that semantic profiling substantially broadens corpus visibility beyond dynamically executable repositories. The semantic fallback recovers tool-level attack-surface information for repositories that cannot be started or exercised by the runtime pipeline.

\begin{table}[!htbp]
\centering
\small
\caption{Semantic capability distribution across MCP tools.}
\label{tab:semantic-full-capabilities}
\begin{tabular}{lr}
\toprule
Capability & Tools \\
\midrule
Network access & 500 \\
Browser automation & 348 \\
Unknown / benign utility & 241 \\
Cloud or SaaS operation & 209 \\
Filesystem access & 204 \\
Code repository operation & 198 \\
Database access & 196 \\
Credential or secret handling & 94 \\
Shell or process execution & 62 \\
\bottomrule
\end{tabular}
\end{table}

Table~\ref{tab:semantic-full-capabilities} summarizes declared capabilities. The most common security-relevant categories involve network, browser, cloud/SaaS, filesystem, code-repository, and database access. Because labels are multi-label, counts do not sum to the number of tools. We interpret these labels as declared attack surface and test-prioritization evidence, not as vulnerability findings.

\subsubsection{Semantic/runtime egress cross-validation.}

Finally, we compare declared egress-related semantics with runtime egress evidence over the repositories that complete targeted exploration. This analysis links semantic profiling to runtime observation, but it is not classifier accuracy against vulnerability ground truth: declared capabilities describe what a server claims it can do, whereas runtime evidence describes what was exercised in one scan scenario.

\begin{table}[!hbtp]
\centering
\small
\setlength{\tabcolsep}{3pt}
\renewcommand{\arraystretch}{1.08}
\caption{Cross-validation between declared egress semantics and runtime network evidence.}
\label{tab:semantic-cross-validation}
\begin{tabular}{@{}p{0.34\columnwidth}r p{0.48\columnwidth}@{}}
\toprule
Relation & Repos & Interpretation \\
\midrule
Declared egress risk and observed egress
  & 5 & Confirmed by runtime evidence \\
Declared egress risk only
  & 14 & Declared capability was not exercised by this run \\
Observed egress only
  & 4 & Potential metadata understatement or classifier miss \\
Neither declared nor observed
  & 10 & No egress evidence in this run \\
\bottomrule
\end{tabular}
\end{table}

Table~\ref{tab:semantic-cross-validation} separates confirmed behavior, unexercised declared capabilities, observed behavior without matching metadata, and runs with neither signal. Declared-only cases are treated as unexercised capabilities rather than false positives, since the relevant behavior may require specific arguments, credentials, services, workflows, or network paths. Observed-only cases suggest metadata understatement, incomplete descriptions, unsupported registration patterns, tool-selection effects, or semantic classifier misses.

Targeted exploration changes this distribution because it invokes more tools and different calls than the shallow scan. We therefore report the targeted comparison as the main deeper-study result, while interpreting it as evidence for one exploration strategy rather than vulnerability ground truth.

\begin{rqbox}
\noindent\textbf{Answer to RQ5.}
Semantic profiling substantially extends visibility beyond dynamic execution. SandScope recovers tool-level metadata for 71 repositories covering 1,127 tools, including 886 with at least one security-sensitive declared capability, and its semantic/runtime egress comparison helps distinguish confirmed behavior, unexercised declared capabilities, and possible metadata understatement.
\end{rqbox}

\subsubsection{Anonymized real-world cases}

To make the aggregate categories in Table~\ref{tab:semantic-cross-validation}
concrete, and to illustrate the additional evidence produced by targeted
exploration, we inspected three dynamically executed repositories from the
corpus.

Repo A demonstrates a source-to-sink witness for environment authority. The
targeted pass selected an environment-inspection tool, seeded a distinct
\texttt{MCP\_ENV\_CANARY\_*} value into the server environment, and observed
the same canary as an exact substring in an MCP \texttt{tools/call} text result.
This case shows that a real MCP implementation can reflect authority-bearing
process state into LLM-visible output under protocol execution.

Repo B demonstrates an input-to-sink witness. The targeted pass generated
schema-valid arguments containing \texttt{MCP\_INPUT\_CANARY\_*} for repository
or query fields of a code-repository tool. The server attempted the requested
operation, failed because the generated identifier was invalid or inaccessible,
and reflected the generated value in MCP-visible error or result text. This
witness is not a vulnerability by itself, but it confirms that SandScope can
observe real client-input-to-result propagation in an MCP server.

Repo C illustrates the ``observed egress only'' category. The invoked
desktop/configuration-oriented tool was not classified as network-capable from
recovered metadata, yet runtime monitoring recorded denied outbound connections
to an external browser-support host. The same repository also produced
MCP-visible local configuration or diagnostic output. We treat this as runtime
observability evidence rather than a source-to-sink witness, because the observed
configuration text was not linked to a seeded source by a byte-level match. This
case shows how metadata can understate runtime behavior when tool descriptions
are incomplete.

We anonymize these repositories because the goal is to validate evidence
categories and runtime observability, not to report repository-specific
vulnerabilities. Together, the cases reinforce SandScope's evidence separation: semantic labels are attack-surface metadata, egress events are execution evidence, and
source-to-sink matches are runtime witnesses for the exercised scenario.

\section{Discussion \& Limitations}
\label{sec:discussion}

SandScope separates runtime evidence from semantic attack-surface metadata and reports only positive, auditable witnesses: source-derived bytes, or normalized/decoded views of them, that appear in LLM-visible MCP sinks. This design makes findings explainable but incomplete. The detector covers exact substrings, suffix-aware snippets, separator-normalized chunking, and a small set of decoder passes, but does not infer arbitrary transformations such as encryption, compression, hashing, semantic rewriting, summarization, or custom encodings. Therefore, absence of a witness is not evidence that no flow exists, and a flow-observed repository is not automatically a vulnerability; it only shows that seeded source bytes reached an LLM-visible sink in the exercised scenario.

Coverage is also limited by what SandScope can execute and observe. Targeted exploration only runs on repositories that complete a shallow MCP session, uses metadata-based tool ranking, avoids destructive operations, and generates minimal schema-guided arguments. It may miss flows requiring authentication, realistic user state, complex workspaces, multi-turn interaction, external services, or side-effectful workflows. Similarly, dynamic scanning is limited by build, startup, and protocol-initialization failures, while source and sink coverage is restricted to modeled environment values, size-capped files, output-surfaced network intents, and observed MCP transcripts. Semantic profiling broadens visibility through declared tool names, descriptions, schemas, registrations, and repository metadata, but these labels indicate attack surface rather than exploitability. SandScope therefore treats semantic metadata as coverage and prioritization evidence, and reserves observed source-to-sink behavior for runtime witnesses.

\section{Related Work}

\paragraph{Black-box taint inference and dynamic analysis.}
SandScope is related to black-box taint inference, which infers input-output dependencies from observable string relationships rather than program instrumentation~\cite{sekar2009efficient,10.1145/1273463.1273490}. However, SandScope is deliberately simpler: it uses sampled source windows, suffix coverage, separator normalization, and a small set of decoder passes to produce auditable witnesses for MCP LLM-visible sinks. It does not implement full approximate taint inference or syntax-aware policy enforcement.

\paragraph{MCP and agent security scanning.}
Recent MCP and agent-security work emphasizes static or orchestration-level defenses. Static-first scanners and agent scans inspect tool metadata, source registrations, dependencies, or configuration risks~\cite{sha2025mcpscan,beurer-kellner2025introducing-mcp-scan,snyk_agent_scan}. Benchmark and system studies further show that tool descriptions, untrusted content, and confused-deputy behavior can induce unsafe tool use without traditional code exploits~\cite{wang2025mcptoxbenchmarktoolpoisoning,embracethered2025mcp,hou2025modelcontextprotocolmcp}. Other defenses reason about workflow separation or toxic capability flows across agent plans~\cite{DBLP:journals/corr/abs-2503-18813,beurer-kellner2025toxic}, and security guidance recommends least privilege and sandboxed MCP deployment~\cite{owasp_genai_mcp_servers_2025}. These approaches are complementary to SandScope: they help identify risky declarations, configurations, or workflows, but do not directly provide cross-implementation runtime witnesses showing that environment, file, or network-intent sources reached LLM-visible MCP outputs.

\paragraph{Sandboxing and secure execution.}

WASM/WASI and OS-level sandboxes provide isolation boundaries for untrusted or semi-trusted code~\cite{10179357,10562203,10.1145/3423211.3425680}. SandScope uses these mechanisms as backends rather than treating isolation as the main contribution. WASI is useful when a portable artifact or shim is available, while native stdio protocol driving preserves fidelity for unmodified MCP servers and should be paired with external OS containment when scanning untrusted native tools. The key distinction is that SandScope adds MCP-aware sink extraction, source-to-sink witness reporting, and semantic capability profiling on top of controlled execution.

Overall, existing work either isolates execution without MCP-aware leakage semantics, or inspects MCP and agent artifacts without dynamic source-to-sink evidence. WASM/WASI runtimes provide containment~\cite{10179357,10562203,10.1145/3423211.3425680}; static MCP and agent scanners inspect metadata, registrations, dependencies, or configuration risks~\cite{sha2025mcpscan,beurer-kellner2025introducing-mcp-scan,snyk_agent_scan}; workflow-level defenses reason about tool-use separation and toxic capability flows~\cite{DBLP:journals/corr/abs-2503-18813,beurer-kellner2025toxic}; and black-box taint analysis infers runtime input-output dependencies outside MCP-specific protocol semantics~\cite{sekar2009efficient}. SandScope complements these approaches by adding MCP-visible sink extraction, auditable runtime witnesses, and semantic capability profiling on top of controlled execution.

\section{Conclusion}

We presented \textbf{SandScope}, an MCP-aware behavioral audit framework for third-party tools in LLM agent supply chains. SandScope combines controlled execution, protocol-aware sink extraction, runtime source-to-sink witness reporting, and semantic capability profiling. It executes portable tools under WASI or drives unmodified stdio MCP servers, extracts LLM-visible sinks from MCP outputs, and reports auditable witnesses when environment, file, or network-intent sources appear in tool results, prompt/messages fields, or structured return payloads.

Our evaluation on controlled cross-language subjects, an evasion benchmark, and a 100-repository MCP corpus shows that SandScope can surface concrete runtime evidence, characterize the limits of lightweight string-level matching, and recover semantic metadata beyond dynamically executable repositories. In the real-world corpus, SandScope completes shallow dynamic scans for 35 repositories and recovers tool-level metadata for 71 repositories covering 1,127 tools, including 886 with at least one security-sensitive declared capability. A targeted exploration pass over the 35 shallow-success repositories successfully re-executes 33 of them and observes source-to-sink witnesses in 12. SandScope is not a complete taint-tracking engine or a standalone vulnerability detector. Its contribution is a practical MCP behavioral audit layer that combines controlled execution, MCP-visible source-to-sink witnesses, targeted exploration, and semantic attack-surface profiling to provide auditable evidence for MCP tool risk.


\clearpage



\begin{thebibliography}{28}


\ifx \showCODEN    \undefined \def \showCODEN     #1{\unskip}     \fi
\ifx \showISBNx    \undefined \def \showISBNx     #1{\unskip}     \fi
\ifx \showISBNxiii \undefined \def \showISBNxiii  #1{\unskip}     \fi
\ifx \showISSN     \undefined \def \showISSN      #1{\unskip}     \fi
\ifx \showLCCN     \undefined \def \showLCCN      #1{\unskip}     \fi
\ifx \shownote     \undefined \def \shownote      #1{#1}          \fi
\ifx \showarticletitle \undefined \def \showarticletitle #1{#1}   \fi
\ifx \showURL      \undefined \def \showURL       {\relax}        \fi
\providecommand\bibfield[2]{#2}
\providecommand\bibinfo[2]{#2}
\providecommand\natexlab[1]{#1}
\providecommand\showeprint[2][]{arXiv:#2}

\bibitem[Alrahis et~al\mbox{.}(2021)]%
        {8836102}
\bibfield{author}{\bibinfo{person}{Lilas Alrahis}, \bibinfo{person}{Muhammad Yasin}, \bibinfo{person}{Nimisha Limaye}, \bibinfo{person}{Hani Saleh}, \bibinfo{person}{Baker Mohammad}, \bibinfo{person}{Mahmoud Al-Qutayri}, {and} \bibinfo{person}{Ozgur Sinanoglu}.} \bibinfo{year}{2021}\natexlab{}.
\newblock \showarticletitle{ScanSAT: Unlocking Static and Dynamic Scan Obfuscation}.
\newblock \bibinfo{journal}{\emph{IEEE Transactions on Emerging Topics in Computing}} \bibinfo{volume}{9}, \bibinfo{number}{4} (\bibinfo{year}{2021}), \bibinfo{pages}{1867--1882}.
\newblock
\href{https://doi.org/10.1109/TETC.2019.2940750}{doi:\nolinkurl{10.1109/TETC.2019.2940750}}


\bibitem[{Anthropic}(2024)]%
        {anthropic2024mcp}
\bibfield{author}{\bibinfo{person}{{Anthropic}}.} \bibinfo{year}{2024}\natexlab{}.
\newblock \bibinfo{booktitle}{\emph{Introducing the Model Context Protocol}}.
\newblock Anthropic.
\newblock
\urldef\tempurl%
\url{https://www.anthropic.com/news/model-context-protocol}
\showURL{%
\tempurl}


\bibitem[Beurer-Kellner and Fischer(2025)]%
        {beurer-kellner2025introducing-mcp-scan}
\bibfield{author}{\bibinfo{person}{Luca Beurer-Kellner} {and} \bibinfo{person}{Marc Fischer}.} \bibinfo{year}{2025}\natexlab{}.
\newblock \bibinfo{booktitle}{\emph{Introducing MCP-Scan: Protecting MCP with Invariant}}.
\newblock
\urldef\tempurl%
\url{https://invariantlabs.ai/blog/introducing-mcp-scan}
\showURL{%
\tempurl}


\bibitem[Beurer-Kellner et~al\mbox{.}(2025)]%
        {beurer-kellner2025toxic}
\bibfield{author}{\bibinfo{person}{Luca Beurer-Kellner}, \bibinfo{person}{Marco Milanta}, {and} \bibinfo{person}{Marc Fischer}.} \bibinfo{year}{2025}\natexlab{}.
\newblock \bibinfo{booktitle}{\emph{Invariant Labs Exposes Novel Prompt Injection Attack Vulnerabilities, “Toxic Flows,” in Agentic Systems \& MCP Servers}}.
\newblock
\urldef\tempurl%
\url{https://invariantlabs.ai/blog/toxic-flow-analysis}
\showURL{%
\tempurl}
\newblock
\shownote{Accessed: 2025-10-06}.


\bibitem[Clause et~al\mbox{.}(2007)]%
        {10.1145/1273463.1273490}
\bibfield{author}{\bibinfo{person}{James Clause}, \bibinfo{person}{Wanchun Li}, {and} \bibinfo{person}{Alessandro Orso}.} \bibinfo{year}{2007}\natexlab{}.
\newblock \showarticletitle{Dytan: a generic dynamic taint analysis framework}. In \bibinfo{booktitle}{\emph{Proceedings of the 2007 International Symposium on Software Testing and Analysis}} (London, United Kingdom) \emph{(\bibinfo{series}{ISSTA '07})}. \bibinfo{publisher}{Association for Computing Machinery}, \bibinfo{address}{New York, NY, USA}, \bibinfo{pages}{196–206}.
\newblock
\showISBNx{9781595937346}
\href{https://doi.org/10.1145/1273463.1273490}{doi:\nolinkurl{10.1145/1273463.1273490}}


\bibitem[Coppolino et~al\mbox{.}(2025)]%
        {10562203}
\bibfield{author}{\bibinfo{person}{Luigi Coppolino}, \bibinfo{person}{Salvatore D'Antonio}, \bibinfo{person}{Giovanni Mazzeo}, \bibinfo{person}{Roberto Nardone}, \bibinfo{person}{Luigi Romano}, {and} \bibinfo{person}{Mathieu Schmitt}.} \bibinfo{year}{2025}\natexlab{}.
\newblock \showarticletitle{WASMBOX: A Lightweight Wasm-Based Runtime for Trustworthy Multi-Tenant Embedded Systems}.
\newblock \bibinfo{journal}{\emph{IEEE Transactions on Emerging Topics in Computing}} \bibinfo{volume}{13}, \bibinfo{number}{2} (\bibinfo{year}{2025}), \bibinfo{pages}{467--480}.
\newblock
\href{https://doi.org/10.1109/TETC.2024.3409817}{doi:\nolinkurl{10.1109/TETC.2024.3409817}}


\bibitem[Debenedetti et~al\mbox{.}(2026)]%
        {debenedetti2026defeating}
\bibfield{author}{\bibinfo{person}{Edoardo Debenedetti}, \bibinfo{person}{Ilia Shumailov}, \bibinfo{person}{Tianqi Fan}, \bibinfo{person}{Jamie Hayes}, \bibinfo{person}{Nicholas Carlini}, \bibinfo{person}{Daniel Fabian}, \bibinfo{person}{Christoph Kern}, \bibinfo{person}{Chongyang Shi}, \bibinfo{person}{Andreas Terzis}, {and} \bibinfo{person}{Florian Tram\`er}.} \bibinfo{year}{2026}\natexlab{}.
\newblock \showarticletitle{Defeating Prompt Injections by Design}. In \bibinfo{booktitle}{\emph{4th IEEE Conference on Secure and Trustworthy Machine Learning}}.
\newblock
\urldef\tempurl%
\url{https://arxiv.org/abs/2503.18813}
\showURL{%
\tempurl}


\bibitem[Debenedetti et~al\mbox{.}(2025)]%
        {DBLP:journals/corr/abs-2503-18813}
\bibfield{author}{\bibinfo{person}{Edoardo Debenedetti}, \bibinfo{person}{Ilia Shumailov}, \bibinfo{person}{Tianqi Fan}, \bibinfo{person}{Jamie Hayes}, \bibinfo{person}{Nicholas Carlini}, \bibinfo{person}{Daniel Fabian}, \bibinfo{person}{Christoph Kern}, \bibinfo{person}{Chongyang Shi}, \bibinfo{person}{Andreas Terzis}, {and} \bibinfo{person}{Florian Tramèr}.} \bibinfo{year}{2025}\natexlab{}.
\newblock \bibinfo{title}{Defeating Prompt Injections by Design}.
\newblock


\bibitem[{Embrace The Red}(2025)]%
        {embracethered2025mcp}
\bibfield{author}{\bibinfo{person}{{Embrace The Red}}.} \bibinfo{year}{2025}\natexlab{}.
\newblock \bibinfo{booktitle}{\emph{MCP: Untrusted Servers and Confused Clients, Plus a Sneaky Exploit}}.
\newblock
\urldef\tempurl%
\url{https://embracethered.com/blog/posts/2025/model-context-protocol-security-risks-and-exploits/}
\showURL{%
\tempurl}
\newblock
\shownote{Accessed: 2025-08-31}.


\bibitem[Gadepalli et~al\mbox{.}(2020)]%
        {10.1145/3423211.3425680}
\bibfield{author}{\bibinfo{person}{Phani~Kishore Gadepalli}, \bibinfo{person}{Sean McBride}, \bibinfo{person}{Gregor Peach}, \bibinfo{person}{Ludmila Cherkasova}, {and} \bibinfo{person}{Gabriel Parmer}.} \bibinfo{year}{2020}\natexlab{}.
\newblock \showarticletitle{Sledge: a Serverless-first, Light-weight Wasm Runtime for the Edge}. In \bibinfo{booktitle}{\emph{Proceedings of the 21st International Middleware Conference}} (Delft, Netherlands) \emph{(\bibinfo{series}{Middleware '20})}. \bibinfo{publisher}{Association for Computing Machinery}, \bibinfo{address}{New York, NY, USA}, \bibinfo{pages}{265–279}.
\newblock
\showISBNx{9781450381536}
\href{https://doi.org/10.1145/3423211.3425680}{doi:\nolinkurl{10.1145/3423211.3425680}}


\bibitem[Hou et~al\mbox{.}(2025)]%
        {hou2025modelcontextprotocolmcp}
\bibfield{author}{\bibinfo{person}{Xinyi Hou}, \bibinfo{person}{Yanjie Zhao}, \bibinfo{person}{Shenao Wang}, {and} \bibinfo{person}{Haoyu Wang}.} \bibinfo{year}{2025}\natexlab{}.
\newblock \bibinfo{title}{Model Context Protocol (MCP): Landscape, Security Threats, and Future Research Directions}.
\newblock
\showeprint[arxiv]{2503.23278}~[cs.CR]
\urldef\tempurl%
\url{https://arxiv.org/abs/2503.23278}
\showURL{%
\tempurl}


\bibitem[Jing et~al\mbox{.}(2025)]%
        {jing-etal-2025-mcip}
\bibfield{author}{\bibinfo{person}{Huihao Jing}, \bibinfo{person}{Haoran Li}, \bibinfo{person}{Wenbin Hu}, \bibinfo{person}{Qi Hu}, \bibinfo{person}{Xu Heli}, \bibinfo{person}{Tianshu Chu}, \bibinfo{person}{Peizhao Hu}, {and} \bibinfo{person}{Yangqiu Song}.} \bibinfo{year}{2025}\natexlab{}.
\newblock \showarticletitle{{MCIP}: Protecting {MCP} Safety via Model Contextual Integrity Protocol}. In \bibinfo{booktitle}{\emph{Proceedings of the 2025 Conference on Empirical Methods in Natural Language Processing}}, \bibfield{editor}{\bibinfo{person}{Christos Christodoulopoulos}, \bibinfo{person}{Tanmoy Chakraborty}, \bibinfo{person}{Carolyn Rose}, {and} \bibinfo{person}{Violet Peng}} (Eds.). \bibinfo{publisher}{Association for Computational Linguistics}, \bibinfo{address}{Suzhou, China}, \bibinfo{pages}{1177--1194}.
\newblock
\showISBNx{979-8-89176-332-6}
\href{https://doi.org/10.18653/v1/2025.emnlp-main.62}{doi:\nolinkurl{10.18653/v1/2025.emnlp-main.62}}


\bibitem[Johnson et~al\mbox{.}(2023)]%
        {10179357}
\bibfield{author}{\bibinfo{person}{Evan Johnson}, \bibinfo{person}{Evan Laufer}, \bibinfo{person}{Zijie Zhao}, \bibinfo{person}{Dan Gohman}, \bibinfo{person}{Shravan Narayan}, \bibinfo{person}{Stefan Savage}, \bibinfo{person}{Deian Stefan}, {and} \bibinfo{person}{Fraser Brown}.} \bibinfo{year}{2023}\natexlab{}.
\newblock \showarticletitle{WaVe: a verifiably secure WebAssembly sandboxing runtime}. In \bibinfo{booktitle}{\emph{2023 IEEE Symposium on Security and Privacy (SP)}}. \bibinfo{pages}{2940--2955}.
\newblock
\href{https://doi.org/10.1109/SP46215.2023.10179357}{doi:\nolinkurl{10.1109/SP46215.2023.10179357}}


\bibitem[Lee et~al\mbox{.}(2025)]%
        {10.1007/978-981-95-4674-9_17}
\bibfield{author}{\bibinfo{person}{Yonghwa Lee}, \bibinfo{person}{Wonseok Choi}, {and} \bibinfo{person}{Donghyun Nam}.} \bibinfo{year}{2025}\natexlab{}.
\newblock \showarticletitle{Supply Chain Threats in the MCP Ecosystem: Attack Vectors and Mitigation Strategies}. In \bibinfo{booktitle}{\emph{Advances in Information and Computer Security: 20th International Workshop on Security, IWSEC 2025, Fukuoka, Japan, November 25–27, 2025, Proceedings}} (Fukuoka, Japan). \bibinfo{publisher}{Springer-Verlag}, \bibinfo{address}{Berlin, Heidelberg}, \bibinfo{pages}{329–349}.
\newblock
\showISBNx{978-981-95-4673-2}
\href{https://doi.org/10.1007/978-981-95-4674-9_17}{doi:\nolinkurl{10.1007/978-981-95-4674-9_17}}


\bibitem[Ntousakis et~al\mbox{.}(2025)]%
        {10.1145/3766882.3767177}
\bibfield{author}{\bibinfo{person}{Grigoris Ntousakis}, \bibinfo{person}{Julian~James Stephen}, \bibinfo{person}{Michael~V. Le}, \bibinfo{person}{Sai Sree~Laya Chukkapalli}, \bibinfo{person}{Teryl Taylor}, \bibinfo{person}{Ian~M. Molloy}, {and} \bibinfo{person}{Frederico Araujo}.} \bibinfo{year}{2025}\natexlab{}.
\newblock \showarticletitle{Securing MCP-based Agent Workflows}. In \bibinfo{booktitle}{\emph{Proceedings of the 4th Workshop on Practical Adoption Challenges of ML for Systems}} (Seoul, Republic of Korea) \emph{(\bibinfo{series}{PACMI '25})}. \bibinfo{publisher}{Association for Computing Machinery}, \bibinfo{address}{New York, NY, USA}, \bibinfo{pages}{50–55}.
\newblock
\showISBNx{9798400722059}


\bibitem[{OWASP GenAI Security Project}(2025)]%
        {owasp_genai_mcp_servers_2025}
\bibfield{author}{\bibinfo{person}{{OWASP GenAI Security Project}}.} \bibinfo{year}{2025}\natexlab{}.
\newblock \bibinfo{title}{CheatSheet -- A Practical Guide for Securely Using Third-Party MCP Servers 1.0}.
\newblock
\urldef\tempurl%
\url{https://genai.owasp.org/resource/cheatsheet-a-practical-guide-for-securely-using-third-party-mcp-servers-1-0/}
\showURL{%
\tempurl}
\newblock
\shownote{Version 1.0; accessed via https://genai.owasp.org/resource/cheatsheet-a-practical-guide-for-securely-using-third-party-mcp-servers-1-0/}.


\bibitem[Prabhakar et~al\mbox{.}(2025)]%
        {NEURIPS2025_5e3661f7}
\bibfield{author}{\bibinfo{person}{Akshara Prabhakar}, \bibinfo{person}{Zuxin Liu}, \bibinfo{person}{Ming Zhu}, \bibinfo{person}{Jianguo Zhang}, \bibinfo{person}{Tulika~Manoj Awalgaonkar}, \bibinfo{person}{Shiyu Wang}, \bibinfo{person}{Zhiwei Liu}, \bibinfo{person}{Haolin Chen}, \bibinfo{person}{Thai Hoang}, \bibinfo{person}{Juan~Carlos Niebles}, \bibinfo{person}{Shelby Heinecke}, \bibinfo{person}{Weiran Yao}, \bibinfo{person}{Huan Wang}, \bibinfo{person}{Silvio Savarese}, {and} \bibinfo{person}{Caiming Xiong}.} \bibinfo{year}{2025}\natexlab{}.
\newblock \showarticletitle{APIGen-MT: Agentic Pipeline for Multi-Turn Data Generation via Simulated Agent-Human Interplay}. In \bibinfo{booktitle}{\emph{Advances in Neural Information Processing Systems}}, \bibfield{editor}{\bibinfo{person}{D.~Belgrave}, \bibinfo{person}{C.~Zhang}, \bibinfo{person}{H.~Lin}, \bibinfo{person}{R.~Pascanu}, \bibinfo{person}{P.~Koniusz}, \bibinfo{person}{M.~Ghassemi}, {and} \bibinfo{person}{N.~Chen}} (Eds.), Vol.~\bibinfo{volume}{38}. \bibinfo{publisher}{Curran Associates, Inc.}
\newblock
\urldef\tempurl%
\url{https://proceedings.neurips.cc/paper_files/paper/2025/file/5e3661f7fe4c8ac5652d62eb3d3c96ea-Paper-Datasets_and_Benchmarks_Track.pdf}
\showURL{%
\tempurl}


\bibitem[Radosevich and Halloran(2025)]%
        {radosevich2025mcpsafetyauditllms}
\bibfield{author}{\bibinfo{person}{Brandon Radosevich} {and} \bibinfo{person}{John Halloran}.} \bibinfo{year}{2025}\natexlab{}.
\newblock \bibinfo{title}{MCP Safety Audit: LLMs with the Model Context Protocol Allow Major Security Exploits}.
\newblock
\showeprint[arxiv]{2504.03767}~[cs.CR]
\urldef\tempurl%
\url{https://arxiv.org/abs/2504.03767}
\showURL{%
\tempurl}


\bibitem[Raheem and Hossain(2025)]%
        {11103638}
\bibfield{author}{\bibinfo{person}{Tayiba Raheem} {and} \bibinfo{person}{Gahangir Hossain}.} \bibinfo{year}{2025}\natexlab{}.
\newblock \showarticletitle{Agentic AI Systems: Opportunities, Challenges, and Trustworthiness}. In \bibinfo{booktitle}{\emph{2025 IEEE International Conference on Electro Information Technology (eIT)}}. \bibinfo{pages}{618--624}.
\newblock
\href{https://doi.org/10.1109/eIT64391.2025.11103638}{doi:\nolinkurl{10.1109/eIT64391.2025.11103638}}


\bibitem[Sekar(2009)]%
        {sekar2009efficient}
\bibfield{author}{\bibinfo{person}{R. Sekar}.} \bibinfo{year}{2009}\natexlab{}.
\newblock \showarticletitle{An Efficient Black-box Technique for Defeating Web Application Attacks}. In \bibinfo{booktitle}{\emph{Proceedings of the 16th Annual Network and Distributed System Security Symposium (NDSS)}}. Internet Society.
\newblock


\bibitem[Sha et~al\mbox{.}(2025)]%
        {sha2025mcpscan}
\bibfield{author}{\bibinfo{person}{Zeyang Sha}, \bibinfo{person}{Hongcheng Li}, \bibinfo{person}{Changhua Chen}, \bibinfo{person}{Run Xiong}, \bibinfo{person}{Shiwen Cui}, \bibinfo{person}{Changhua Meng}, {and} \bibinfo{person}{Weiqiang Wang}.} \bibinfo{year}{2025}\natexlab{}.
\newblock \bibinfo{title}{MCPSCAN}.
\newblock
\urldef\tempurl%
\url{https://github.com/antgroup/Trustworthy_LM/mcp-scan}
\showURL{%
\tempurl}


\bibitem[{Snyk}(2026)]%
        {snyk_agent_scan}
\bibfield{author}{\bibinfo{person}{{Snyk}}.} \bibinfo{year}{2026}\natexlab{}.
\newblock \bibinfo{title}{Agent Scan: Security scanner for AI agents, MCP servers and agent skills}.
\newblock \bibinfo{howpublished}{\url{https://github.com/snyk/agent-scan}}.
\newblock
\newblock
\shownote{Accessed: 2026-06-19}.


\bibitem[Song et~al\mbox{.}(2025)]%
        {song2025protocolunveilingattackvectors}
\bibfield{author}{\bibinfo{person}{Hao Song}, \bibinfo{person}{Yiming Shen}, \bibinfo{person}{Wenxuan Luo}, \bibinfo{person}{Leixin Guo}, \bibinfo{person}{Ting Chen}, \bibinfo{person}{Jiashui Wang}, \bibinfo{person}{Beibei Li}, \bibinfo{person}{Xiaosong Zhang}, {and} \bibinfo{person}{Jiachi Chen}.} \bibinfo{year}{2025}\natexlab{}.
\newblock \bibinfo{title}{Beyond the Protocol: Unveiling Attack Vectors in the Model Context Protocol (MCP) Ecosystem}.
\newblock
\showeprint[arxiv]{2506.02040}~[cs.CR]
\urldef\tempurl%
\url{https://arxiv.org/abs/2506.02040}
\showURL{%
\tempurl}


\bibitem[Wang et~al\mbox{.}(2025)]%
        {wang2025mcptoxbenchmarktoolpoisoning}
\bibfield{author}{\bibinfo{person}{Zhiqiang Wang}, \bibinfo{person}{Yichao Gao}, \bibinfo{person}{Yanting Wang}, \bibinfo{person}{Suyuan Liu}, \bibinfo{person}{Haifeng Sun}, \bibinfo{person}{Haoran Cheng}, \bibinfo{person}{Guanquan Shi}, \bibinfo{person}{Haohua Du}, {and} \bibinfo{person}{Xiangyang Li}.} \bibinfo{year}{2025}\natexlab{}.
\newblock \bibinfo{title}{MCPTox: A Benchmark for Tool Poisoning Attack on Real-World MCP Servers}.
\newblock
\showeprint[arxiv]{2508.14925}~[cs.CR]
\urldef\tempurl%
\url{https://arxiv.org/abs/2508.14925}
\showURL{%
\tempurl}


\bibitem[Xu et~al\mbox{.}(2017)]%
        {7958617}
\bibfield{author}{\bibinfo{person}{Dongpeng Xu}, \bibinfo{person}{Jiang Ming}, {and} \bibinfo{person}{Dinghao Wu}.} \bibinfo{year}{2017}\natexlab{}.
\newblock \showarticletitle{Cryptographic Function Detection in Obfuscated Binaries via Bit-Precise Symbolic Loop Mapping}. In \bibinfo{booktitle}{\emph{2017 IEEE Symposium on Security and Privacy (SP)}}. \bibinfo{pages}{921--937}.
\newblock
\href{https://doi.org/10.1109/SP.2017.56}{doi:\nolinkurl{10.1109/SP.2017.56}}


\bibitem[Zhang et~al\mbox{.}(2026)]%
        {zhang2026mcp}
\bibfield{author}{\bibinfo{person}{Dongsen Zhang}, \bibinfo{person}{Zekun Li}, \bibinfo{person}{Xu Luo}, \bibinfo{person}{Xuannan Liu}, \bibinfo{person}{Pei~Pei Li}, {and} \bibinfo{person}{Wenjun Xu}.} \bibinfo{year}{2026}\natexlab{}.
\newblock \showarticletitle{{MCP} Security Bench ({MSB}): Benchmarking Attacks Against Model Context Protocol in {LLM} Agents}. In \bibinfo{booktitle}{\emph{The Fourteenth International Conference on Learning Representations}}.
\newblock
\urldef\tempurl%
\url{https://openreview.net/forum?id=irxxkFMrry}
\showURL{%
\tempurl}

\bibitem[Zhao et~al\mbox{.}(2025)]%
        {Zhao2025ParasitesIT}
\bibfield{author}{\bibinfo{person}{Shuli Zhao}, \bibinfo{person}{Qinsheng Hou}, \bibinfo{person}{Zihan Zhan}, \bibinfo{person}{Yanhao Wang}, \bibinfo{person}{Yuchong Xie}, \bibinfo{person}{Yu Guo}, \bibinfo{person}{Libo Chen}, \bibinfo{person}{Shenghong Li}, {and} \bibinfo{person}{Zhi Xue}.} \bibinfo{year}{2025}\natexlab{}.
\newblock \showarticletitle{Parasites in the Toolchain: A Large-Scale Analysis of Attacks on the MCP Ecosystem}.
\newblock
\urldef\tempurl%
\url{https://api.semanticscholar.org/CorpusID:281203664}
\showURL{%
\tempurl}


\bibitem[Zhao et~al\mbox{.}(2026)]%
        {zhao2026proactivedefensellmjailbreak}
\bibfield{author}{\bibinfo{person}{Weiliang Zhao}, \bibinfo{person}{Jinjun Peng}, \bibinfo{person}{Daniel Ben-Levi}, \bibinfo{person}{Zhou Yu}, {and} \bibinfo{person}{Junfeng Yang}.} \bibinfo{year}{2026}\natexlab{}.
\newblock \bibinfo{title}{Proactive defense against LLM Jailbreak}.
\newblock
\showeprint[arxiv]{2510.05052}~[cs.CR]
\urldef\tempurl%
\url{https://arxiv.org/abs/2510.05052}
\showURL{%
\tempurl}

\end{thebibliography}
\end{document}